\documentclass[traditabstract]{aa} 
\usepackage{graphicx}
\usepackage{txfonts}
\usepackage{amssymb}
\usepackage{natbib}

\begin{document}
   \title{Cluster galaxies in XMMU J2235-2557: galaxy population
 properties in most massive environments at
 $z\sim1.4$\thanks{Based on observations made with the NASA/ESA Hubble
          Space Telescope under Program IDs 10698, 10496, and 10531,
          and with the Very Large Telescope at the ESO Paranal
          Observatory under Program IDs 60.A-9284, 072.A-0706,
          073.A-0737, 074.A-0023, 077.A-0177, 077.A-0110, 079.A-0758,
          and 081.A-0759}}


   \author{V.  Strazzullo
          \inst{1}
          \and P.  Rosati \inst{2} \and  M.
Pannella \inst{1} \and  R.  Gobat \inst{3} \and 
J. S. Santos \inst{4} \and  M. Nonino \inst{4} \and 
R. Demarco \inst{5} \and 
C. Lidman \inst{6} \and 
 M. Tanaka \inst{2,7} \and \\
C. R. Mullis \inst{8} \and  C. Nu\~nez \inst{2} \and 
A. Rettura \inst{9} \and  M. J. Jee \inst{10} \and 
H. B\"ohringer \inst{11} \and  R. Bender \inst{11,12} \and 
R. J. Bouwens \inst{13,14} \and 
K. Dawson \inst{15} \and\\
R. Fassbender \inst{11} \and  M. Franx \inst{13} \and  S. 
Perlmutter \inst{16} \and  M. Postman \inst{17}}

\institute{
National Radio Astronomy Observatory, 1003 Lopezville  Rd., Socorro, NM 87801, USA  \email{vstrazzu@nrao.edu}
\and
European Southern Observatory, Karl Schwarzschild  Strasse 2, 85748 Garching bei Muenchen, Germany
\and
CEA, Laboratoire AIM-CNRS-Universit\'e Paris Diderot, Irfu/SAp, Orme des Merisiers, 91191 Gif-sur-Yvette, France
\and
INAF-Osservatorio Astronomico di Trieste, via Tiepolo 11, 34131 Trieste, Italy
\and
Department of Astronomy, Universidad de Concepci{\'o}n. Casilla 160-C, Concepci\'on, Chile
\and
Australian Astronomical Observatory, PO Box 296, Epping, NSW 1710, Australia
\and
Institute for the Physics and Mathematics of the Universe, The
University of Tokyo,  5-1-5 Kashiwanoha, Kashiwa-shi, Chiba 277-8583, Japan
\and
Wells Fargo Bank, 4525 Sharon Road, Charlotte, NC 28211, USA
\and
Department of Physics and Astronomy, University of California,
Riverside, CA 92521, USA
\and
Department of Physics, University of California, Davis, One Shields Avenue, Davis, CA 95616, USA
\and
Max-Planck-Institut f\"ur Extraterrestrische Physik,
  Giessenbachstrasse, 85748 Garching, Germany
\and
Universit{\"a}ts-Sternwarte, Scheinerstrasse 1, Munich D-81679,  Germany
\and
Leiden Observatory, Leiden University, P.O. Box 9513, 2300  RA Leiden, Netherlands
\and
Astronomy Department, University of California,
  Santa Cruz, CA 95064, USA
\and
Department of Physics and Astronomy, University of Utah, Salt Lake City, UT 84112, USA
\and
Lawrence Berkeley National Laboratory, 1 Cyclotron 
Rd., Berkeley, CA 94720, USA
\and
Space Telescope Science Institute, 3700 San Martin 
Drive, Baltimore, MD 21218, USA }

   \date{Received ; accepted }

 
  \abstract
{ We present a multi--wavelength study of galaxy populations in the core
of the massive, X--ray luminous cluster XMMU J2235 at z=1.39, based on
high quality VLT and HST photometry at optical and near--infrared
wavelengths.

We derive luminosity functions in the z, H, and K$_{s}$ bands,
approximately corresponding to restframe U, R and z band. These show a
faint--end slope consistent with being flat, and a characteristic
magnitude M$^*$ close to passive evolution predictions of M$^*$ of
local massive clusters, with a formation redshift $z>2$.

The color--magnitude and color--mass diagrams show evidence of a tight
red sequence (intrinsic scatter $\lesssim0.08$) of massive galaxies
already in place, with overall old stellar populations and generally
early--type morphology. Beside the red colors, these massive
($>6 \cdot 10^{10}$M$_{\odot}$) galaxies typically show early-type
spectral features, and rest-frame far-UV emission consistent with very
low star formation rates (SFR$<0.2$M$_{\odot}$/yr). 

Star forming spectroscopic members, with SFR of up to
$\sim$100M$_{\odot}$/yr, are all located at clustercentric distances
$\gtrsim$250kpc, with the central cluster region already appearing
effectively quenched. Most part of the cluster galaxies more massive
than $6\cdot10^{10}$M$_{\odot}$ within the studied area do not appear
to host significant levels of star formation.

The high--mass end galaxy populations in the core of this
cluster appear to be in a very advanced evolutionary stage, not only
in terms of formation of the stellar populations, but also of the
assembly of the stellar mass. The high-mass end of the galaxy stellar
mass function is essentially already in place.  The stellar mass
fraction estimated within $r_{500}$ ($\sim1\%$, Kroupa IMF) is already
similar to that of local massive clusters.

On the other hand, surface brightness distribution modeling of the
massive red sequence galaxies may suggest that their size is often
smaller than expected based on the local stellar mass vs size
relation. An evolution of the stellar mass vs size relation might
imply that, in spite of the overall early assembly of these sources,
their evolution is not complete, and processes like minor (and likely
dry) merging might still shape the structural properties of these
objects to resemble those of their local counterparts, without
substantially affecting their stellar mass or host stellar
populations. Nonetheless, a definite conclusion on the actual
relevance of size evolution for the studied early-type sample is
precluded by possible systematics and biases.}


\titlerunning{Cluster galaxies at z$\sim$1.4}
\authorrunning{Strazzullo et al.}

   \maketitle

\section{Introduction}

Galaxy clusters at high redshift provide us with a unique yet rare
chance to investigate the effect of the highest density environments
on the evolution of galaxy populations. Due to the extreme rarity of
massive galaxy clusters, and especially more so at early cosmic
epochs, even wide--area deep surveys do not probe these peculiar
environments which thus need to be searched and identified in
specifically designed surveys, and then followed--up with observations
across a wide range of wavelengths in order to maximize the scientific
return of their discovery.

Thanks to considerable efforts with on-going cluster surveys, which
are utilizing a variety of methods \citep[e.g.,][and references
therein]{gladders2000,rosati2002,eisenhardt2004,wilson2005,wittman2006,fassbender2008,staniszewski2009,andreon2009,demarco2010},
the number of high-redshift clusters is increasing, however only a
handful of spectroscopically confirmed galaxy clusters beyond redshift
1.3 have been discovered to date
\citep{mullis2005,stanford2005,stanford2006,eisenhardt2008,wilson2009,papovich2010,tanaka2010}.

Galaxy clusters at such high redshifts are not only important for
studying the emergence of the large scale structure and for
constraining cosmological models, but also specifically for the
evolution of massive early--type galaxies which seem to dominate 
massive cluster galaxy populations even beyond redshift one. The
further back in cosmic time we can reach, and thus the closer we can
get to the major stages of the galaxy formation process, the tighter
are the constraints we can set on the evolution of these galaxies, as
well as on the relevance of the cluster environment in shaping their
physical properties \citep[among many others,][and references
therein]{toft2004,blakeslee2006,andreon2006,strazzullo2006,depropris2007,delucia2007,kodama2007,holden2007,zirm2008,andreon2008,lidman2008,gobat2008,muzzin2008,rettura2010,mei2009,collins2009,hilton2009,rosati2009}.

While the detailed study of galaxies with stellar masses $M\lesssim
10^{10} M_\odot$ is still limited at $z>1$, even with the current
generation of 10m class telescopes, massive galaxy populations can be
studied out to high redshift in sufficient detail, in terms of their
structure, star formation histories, and stellar masses. These studies
allow us to constrain the evolution of massive galaxies in more
nuances than the old, and now widely superseded in its original form,
`` monolithic vs hierarchical'' question. This provides valuable input
to theoretical modeling of galaxy evolution, the detailed comparison
of cluster and field galaxies being one of the many examples
\citep{menci2008,romeo2008}.

Taking advantage of the full range of available observations, across
the widest possible range of cosmic epochs and environments, has
become particularly important in order to probe different aspects of
galaxy evolution, and to test specific predictions of theoretical
models. As an example, the introduction of various forms of so-called
feedback mechanisms to modulate galaxy evolution has reconciled
previously considered ``anti--hierarchical'' observations with
hierarchical predictions \citep[e.g.,][and references
therein]{delucia2006,bower2006}, and elucidated the decoupling between
star formation and mass assembly histories of massive galaxies. In
order to probe these two processes independently, galaxy evolution has
to be studied by directly sampling galaxy populations up to the
highest redshifts.

In this work we present multi--wavelength observations of the X--ray
luminous galaxy cluster XMMU J2235-2557 (hereafter XMMU J2235, 
RA= $22^h35^m20^s.82$, Dec=$-25\degr57'40.3''$, J2000) at $z=1.39$,
which is the most massive virialized structure discovered beyond
redshift one \citep{jee2009,rosati2009}, spectroscopically confirmed
with 30 cluster members with redshift in the range $1.37<z<1.41$. This
is a follow-up work of the first multi-wavelegth analysis presented in
\citet{rosati2009}, which uses an enhanced data set and is aimed at
obtaining a more comprehensive picture on the star formation, stellar
mass distribution, and morphological structure of the galaxy
populations in the cluster central regions.

Throughout this paper, we adopt a $H_0$=70 km s$^{-1}$ Mpc$^{-1}$,
$\Omega_M$=0.3, $\Omega_\Lambda$=0.7 cosmology, and the AB magnitude system.

\begin{figure}[t!]
\vspace{.5cm}
\centering
\includegraphics[width=8.7cm]{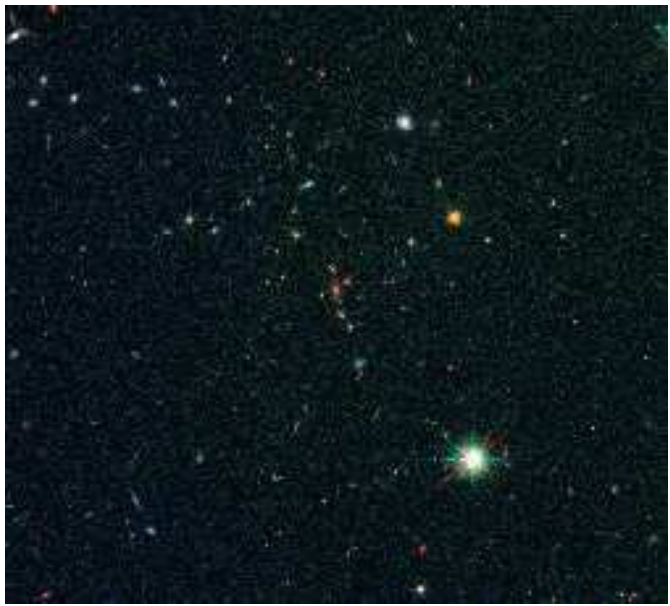}
\vspace{.2cm}
\caption{A color image of XMMU J2235 obtained by combining the ACS
  F775W and F850LP and NICMOS F160W images (North is up, East to the
  left). The image shows a region of about $2\times2$ arcmin$^2$,
  corresponding to about $1\times1$ Mpc$^2$ at the cluster redshift. [Image resolution significantly degraded for astro-ph submission]
\label{fig:colorimage}}
\end{figure}

\section{Data and derived quantities}

\subsection{Observations and catalog production}

This work is based on a multi--wavelength dataset collected in the
field of XMMU J2235 with the VLT and HST in the wavelength range
3500\AA~ to 2$\mu$m. In the following we use VLT photometry (VIMOS U,
FORS2 R, HAWK-I J and K$_{s}$) and spectroscopy (FORS2), and HST
photometry (ACS F775W and F850LP -- hereafter i and z, and NICMOS
F160W -- hereafter H). A description of the first available HST/ACS
data and VLT spectroscopic campaign has been published in
\citet{rosati2009}; in this work we use the full presently
available ACS dataset, including images from program GO-10496
\citep{dawson2009}, as was presented in \citet{jee2009}.  VLT/HAWK-I
data have been published in \citet{lidman2008}, while VLT/FORS2 data
were presented in \citet{mullis2005}, thus we refer to these papers
for a full description of these observations.

The NICMOS F160W data we used in this work were obtained in August
2008 (GO 14, Proposal ID 10531, PI: C. Mullis), resulting in a mosaic
of about 2.5$\times$2.5 arcmin$^2$ with an average exposure time of
$\sim1$hr. 

The U band data (ESO 079.A-0758, PI: M. Nonino) were obtained in
August 2007 with VIMOS at the VLT. 32 dithered observations were
collected for a total of 6hrs 20m with seeing conditions ranging from
0".5 to 1".2.  The images were reduced and stacked in a similar
fashion to \citet{nonino2009}. The zero point of the mosaic was
derived using standard stars observed in photometric nights.  The
limiting magnitude of the mosaic covering the cluster was estimated
from the counts distribution in a 2'' aperture\footnote{Here and in
the following, aperture sizes refer to the aperture diameter.}
centered on 5000 random points. After correcting for aperture effects
(0.19 mag), and for the effect of the correlation in the noise
introduced by the coaddition step (0.2 mags), the final value is 28.8
AB mag ($1\sigma$).

\begin{table}[h]
\caption{Summary of the imaging data used in this work. Columns 1 and
  2 list the instruments and filters with which images were
  acquired. Column 3 gives an estimate of the image
  resolution. Column 4 gives the total area of the image/mosaic in
  each passband, and the actual area used to build the
  multi--wavelength catalog. Further details about these images, and
  additional information specific to their use for different purposes
  in this work, are given in the text.}
\label{tab:data}
\centering   
\begin{tabular}{c c c c}     
\hline\hline                 
  \vspace{-0.1cm}  \\
Telescope/Instrument & Filter & FWHM & Area (total/used)\\  
 &  & (arcsec) & (arcmin$^2$)\\   
\hline                       
  \vspace{-0.1cm}  \\
VLT/VIMOS & U & 0.8 & 50/10.3 \vspace{0.06cm} \\  
VLT/FORS2 & R$\_$SPECIAL &  0.75  & 50/10.3 \vspace{0.06cm} \\
HST/ACS   & F775W &  0.1 & 11/10.3  \vspace{0.06cm} \\
HST/ACS   & F850LP  & 0.1 & 11/10.3  \vspace{0.06cm} \\
VLT/HAWK-I & J &   0.55  & 180/10.3 \vspace{0.06cm} \\ 
HST/NICMOS & F160W & 0.35 & 6.2/6.2  \vspace{0.06cm} \\
VLT/HAWK-I & K$_s$ & 0.4 & 180/10.3  \vspace{0.06cm} \\
\hline                       
\end{tabular}
\end{table}

Source extraction and photometry were performed with SExtractor
\citep{sextractor}, either in single--image (for the determination of
MAG\_AUTOs used for deriving the luminosity functions) or in
dual--image mode (for the determination of aperture colors used for
deriving the broad--band SEDs).  We adopted MAG\_AUTO as an estimate
of total magnitude, while colors (and thus the broad--band SEDs) were
estimated in apertures of 1'' and 1.5''. These aperture sizes
where chosen based on the typical angular size of galaxies at the
redshift we are interested in, and on the resolution of the available
data. The 1'' apertures allow us to measure colors of cluster members
within 1-2 effective radii (see section \ref{sec:stelpopcmds}), while
for estimating stellar masses we used the 1.5'' apertures as a
compromise between optimising the S/N, minimising the effect of
neighbors, and reducing the errors on the correction to total masses
(see section \ref{sec:Mstars}).

The angular resolution of our imaging data ranges from a FWHM of
$\sim$0.1'' for the ACS images to $\sim$0.8'' for ground-based optical
images (see Table \ref{tab:data}). In order to match these different
resolutions in the multicolor photometric catalog, aperture
corrections were applied in each passband as estimated from the
growth--curve of point--like sources in the field \citep[similar to
the approach described in e.g.][]{rettura2006}.  This
resolution-matching approach allows us to measure accurate photometry
without degrading the image quality in any of the passbands, 
which is an important advantage when dealing with a crowded field such
as a cluster core\footnote{For aperture sizes used in this work, and
for our worse resolution of $\sim0.8''$, we estimate through
simulation of synthetic sources (with the IRAF task mkobjects) that
this resolution-matching approach corrects for systematics of 0.1 to
0.5 mag (depending on the aperture size), with a residual systematic
offset of 0.01-0.04 mag for a de Vaucouleurs profile with effective
radius 0.24" ($\sim$2kpc at z=1.39, about the median effective radius of
our red-sequence sample), and $\sim$0.05 mag for effective radii of 1.2"
($\sim$10kpc at z=1.39, about our largest effective radius), and a scatter
of about 0.01-0.02 mag.}.

Magnitudes were corrected for
Galactic extinction according to \citet{schlegel1998}.

Multi-wavelength photometric and morphological catalogs for the
available sample of $\sim30$ spectroscopically confirmed cluster
members will be published in a forthcoming paper (Nu\~nez et al., in
preparation).

\subsection{Photometric redshifts}

We used the U,R,i,z,J,H,K$_{s}$ photometric coverage of the XMMU J2235
field, to estimate photometric redshifts (photo--zs) by comparing the
observed photometry with a library of 33 SED templates covering a
range of star-formation histories, ages and dust content. Together
with local galaxy templates
\citep[e.g.,][]{cww,mannuccitemplates,kinney1996}, we used a set of
semi--empirical templates based on observations plus fitted SED models
\citep{maraston1998,bc03} of $\sim 300$ galaxies in the FORS Deep
Field \citep{heidt2003} and Hubble Deep Field \citep{hdfwilliams1996},
in order to better represent objects to higher redshifts.  A different
prior on the redshift distribution is assumed for different types of
templates (e.g. an old local elliptical template is assumed to be
increasingly unlikely at higher redshifts, while templates
corresponding to young stellar populations or QSOs are assumed to have
a basically flat likelihood across the whole redshift range
probed). In addition, a weak broad prior on the absolute optical and
NIR magnitude lowers the probability to have magnitudes brighter than
-25 and fainter than -13.  Ly-$\alpha$ forest depletion of galaxy
templates is implemented according to \citet{madau1995}. The
``best--fit'' photo--z $z_{phot}$ is chosen as the redshift maximizing
the probability among all templates, and an error on $z_{phot}$ is
defined as $e_{zphot}=[\Sigma_{ij}(z_{i}-z_{phot})^{2} P_{ij}]^{1/2}$,
with $z_{i}$ the considered redshift steps, and $P_{ij}$ the
contribution of the j-th template to the total probability function at
redshift $z_{i}$. We refer to
\citet{bender2001,gabasch2004,brimioulle2008} for a detailed
description of both the photo--z estimation method, and the
construction of the templates.

Systematic offsets between the measured and template--predicted colors
as a function of redshift (which may be due for instance to errors in
the estimated zero--point and aperture corrections, uncertainties in
the filter response curves, or systematics in the templates) were
estimated using $\sim70$ spectroscopic redshifts available within the
$\sim 3\arcmin \times 3\arcmin$ field, and corrected for in order to
minimize the systematic shift between observed and predicted colors
for well--fitted spectroscopic galaxies.

\begin{figure}[t!]
\centering
\hspace{-.3cm}
\includegraphics[width=8.9cm, clip, bb= 134 320 438 750]{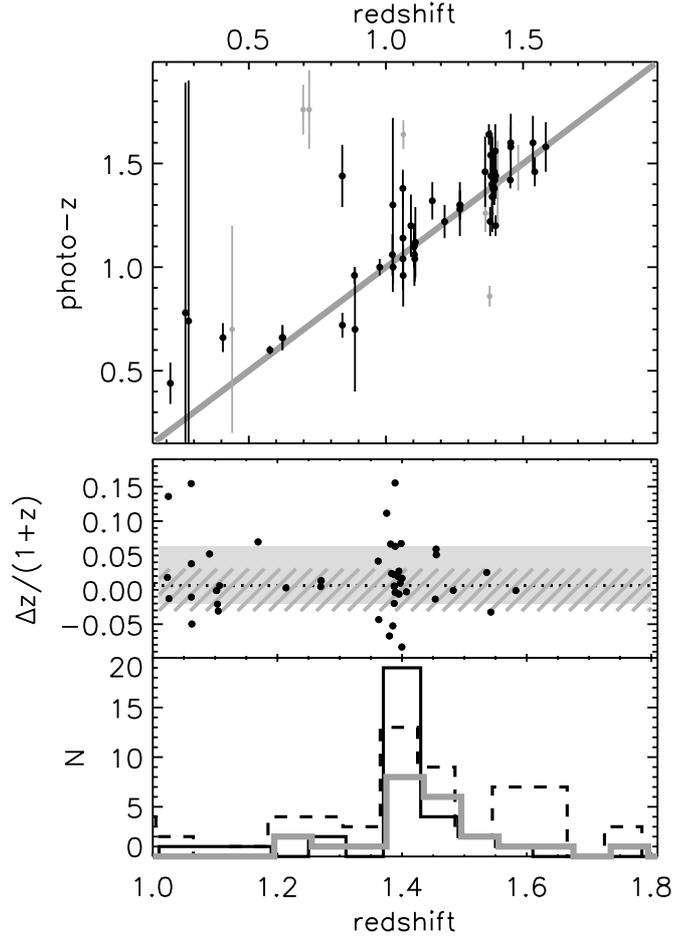}
\caption{Top panel: photometric vs spectroscopic redshifts for the
  whole spectroscopic sample within the considered field (one point at
  z=3.26 with photo-z=3.44 is not plotted). The gray line traces the
  bisector.  Gray symbols show uncertain spectroscopic redshifts
  (quality flag B or worse, see \citet{rosati2009}; these objects are
  plotted together with the secure redshifts in the lower panels).
  Middle panel: $\Delta$z/(1+z) as a function of redshift for the
  spectroscopic sample at $1<z<2$ (one catastrophic outlier with an
  (uncertain) redshift z=1.06 and a photo-z=1.64 falls out of the
  plot). The dotted line and the gray shaded area show the median and
  the $16^{th}-84^{th}$ percentile range of $\Delta z/(1+z)$. The
  hatched area shows the $\Delta z/(1+z)$ scatter as estimated by the
  NMAD estimator.  Bottom panel: the spectroscopic (solid black line)
  and photometric (dashed line) redshift distribution of bright
  (H$<22.2$) galaxies in the studied field. The gray histogram shows
  instead the photometric redshift distribution of all spectroscopic
  cluster members. Histograms are slightly offset for clarity.
\label{fig:zz}}
\end{figure}

The photo-z performance against the available spectroscopic sample is
shown in the top panel of Figure \ref{fig:zz}. The figure clearly
shows the impact of the lack of coverage between the U and R passbands
at lower redshifts, in particular at z$\lesssim$0.6 where the 4000\AA~
break is poorly sampled by the available photometric coverage, often
resulting in unconstrained, inaccurate photo-z's.  However, by
comparison with a spectroscopic sample of $\sim$50 objects in the
redshift range $1<z<2$ ($\sim40$\% of these are cluster members), we
estimate a systematic offset (median $\Delta$z/(1+z)) of 0.006, a
scatter of $\Delta$z/(1+z)$< 6\%$\footnote{The NMAD estimator
\citep{hoaglin1983,ilbert2009} gives a scatter of 0.045 for the whole
spectroscopic sample, and 0.03 in the range $1<z<2$.}, and 4\%
catastrophic outliers (two objects with $|\Delta$z/(1+z)$|\geq20\%$;
we note that both sources have a spectroscopic redshift deemed
uncertain, with a quality flag B or worse, see
\citet{rosati2009}). The middle panel of Figure \ref{fig:zz} shows in
more detail the photo-z performance in the $1<z<2$ range, suggesting
that in the relevant redshift range we can derive accurate enough
photo--zs\footnote{Out of 7 catastrophic failures over the whole
$0<z<3.5$ redshift range, four have an uncertain spectroscopic
redshift, and two are z$\sim$0.3 sources for which the photo-z is
essentially unconstrained, as the errorbars in Figure \ref{fig:zz}
show.}. In the bottom panel of Figure \ref{fig:zz} we compare the
spectroscopic (solid black line) and photometric (dashed line)
redshift distributions in this redshift range for bright galaxies
($<$M$^*$+1), where we have enough spectroscopic coverage (see below,
also figures in section \ref{sec:stelpopcmds}). While the peak at the
cluster redshift is also visible in the photo--z distribution, the
gray histogram shows the spread in photo--zs of all available
spectroscopic members (regardless of their magnitude).

By comparison with the available spectroscopic sample, we estimate
that by selecting photo-z candidates within 3$\sigma$ of the cluster
redshift ($1<z<1.8$) we include all spectroscopic members, while
slightly more than half of the selected sources are interlopers (94\%
of which in the range $1<z_{spec}<1.8$)\footnote{While the 3$\sigma$
photo-z retained samples are used to include all plausible cluster
members, for the purpose of statistical analysis (as for instance in
section \ref{sec:MF}), in the following we also use 2$\sigma$ photo-z
retained samples. Based on the comparison with the spectroscopic
sample, these too are estimated to be virtually 100\% complete, and
affected by an almost 50\% contamination by interlopers. In the
stellar mass range relevant to this work (M$>10^{10}$M$_{\odot}$), the
difference between the two samples is just $\sim10$ sources, with
almost 90\% of the 3$\sigma$ sample belonging to the 2$\sigma$ sample
as well.}.

\subsection{Stellar masses}
\label{sec:Mstars}

Stellar masses were obtained by fitting \citet{bc03} models to the
(resolution--matched) U,R,i,z,J,H,K$_{s}$ SEDs as measured in 1.5''
apertures \citep[as described in e.g.,
][]{gobat2008,rosati2009}. Stellar masses are determined for models of
fixed solar metallicity and with a \citet{salpeter1955} IMF, however for the
purpose of comparison with other literature samples (see below)
stellar masses are converted to \citet{kroupa2001} IMF
masses\footnote{Log(Mass$_{Kroupa}$) = Log(Mass$_{Salpeter}$) -
0.19.}. Different determinations of the stellar masses were derived,
using different star formation histories (exponentially declining,
delayed exponentially declining, with or without a recent burst of
star formation), including or not dust attenuation. These different
determinations mostly agree within better than 50\%, (the scatter
between different determinations is about $\sim$30\%), and never
differ by more than a factor of two.  We will consider this $\sim30\%$
scatter as an estimate of the typical error on the stellar masses due
to different SFHs and dust properties, however we remind the reader
that this estimate does not include other systematics, including those
due to IMF and metallicity being different from what we assumed.

In order to obtain total stellar masses, the masses derived from the
SEDs measured within a radius of 0.75'' were renormalized by the flux
ratio between the 1.5'' aperture magnitude and the total (MAG\_AUTO)
magnitude in the z band.  This approximation neglects the effects of a
change in stellar mass--to--light ratio in the very external regions
of the galaxies. However, given the typical galaxy sizes, the
correction is generally small: among the sample relevant to this work
($1<$z$<2$, log(M$_{*}$/M$_{\odot}$)$>$10.4), the median correction is
less than 10\%, and for the great part of the sources is less than
50\%.  We thus expect this error not to significantly affect our
results, also in view of the uncertainties generally affecting
SED-determined stellar masses \citep[e.g.,][]{longhetti2009}.

\subsection{Structural properties}
\label{sec:galfit}

We used the GALFIT \citep{peng2002} software to model surface
brightness profiles of sources brighter than $z\sim24.2$. We set this
magnitude limit based on S/N considerations, according to previous
results on the accuracy of the retrieved parameters as a function of
magnitude (S/N) from simulations of surface brightness fitting
\citep[e.g., among others,][see also below]{ravindranath2006,pannella2009b}, as
well as visual morphology \citep{postman2005}). This is sufficient to
probe the morphologies of red galaxies in XMMU J2235 down to
$\sim$M$^{*}$+1.

We fitted PSF-convolved \citet{sersic1968} profiles to ACS z band
 images, using the PSF models derived from principal component
 analysis \citep{jee2007,jee2009}; using a stack of high S/N,
 unsaturated point-like sources in the field does not change our
 results.  The possibility offered by GALFIT to simultaneously fit
 multiple sources is particularly useful in a cluster environment:
 when fitting a galaxy surface brightness, all nearby
 sources\footnote{Sources within a region of dimensions about 4 times
 the dimension of the source as estimated through SExtractor
 parameters (ISO\_AREA, position angle).} brighter than
 mag$_{\textrm{galaxy}}$+3 were also modeled at the same time,
 reducing biases produced by contamination by neighbors.

A detailed description of the morphological properties of bright
cluster galaxies, and a comprehensive morphological analysis including
the high--resolution ground--based NIR data, will be presented in
Nu\~nez et al. (in preparation). In the following we only use the Sersic
index $n_{Sersic}$ to broadly classify galaxies in early and late
morphological types, and the estimated (circularized) effective radius
as an estimate of the galaxy size.

\section{Luminosity functions}
\label{sec:LFs}

The luminosity function (LF) of galaxies in the central region of XMMU J2235
was determined in three passbands (z, H, K$_{s}$), roughly
corresponding to restframe U, R, and z bands. The  regions
actually used for the LF determination were about 7.5, 6, and 7
arcmin$^2$, in the z, H, and K$_{s}$ bands, respectively. At the
cluster redshift, the area probed extends out to a clustercentric
distance of about 700 kpc. The completeness limits of the z, H, and
K$_{s}$ band images was estimated based on the turnover of the number
counts of sources with $S/N>10$, and turn out to be $\sim$ 25.3, 25,
and 23, respectively.

\begin{figure*}[ht!]
\centering
\includegraphics[width=\textwidth]{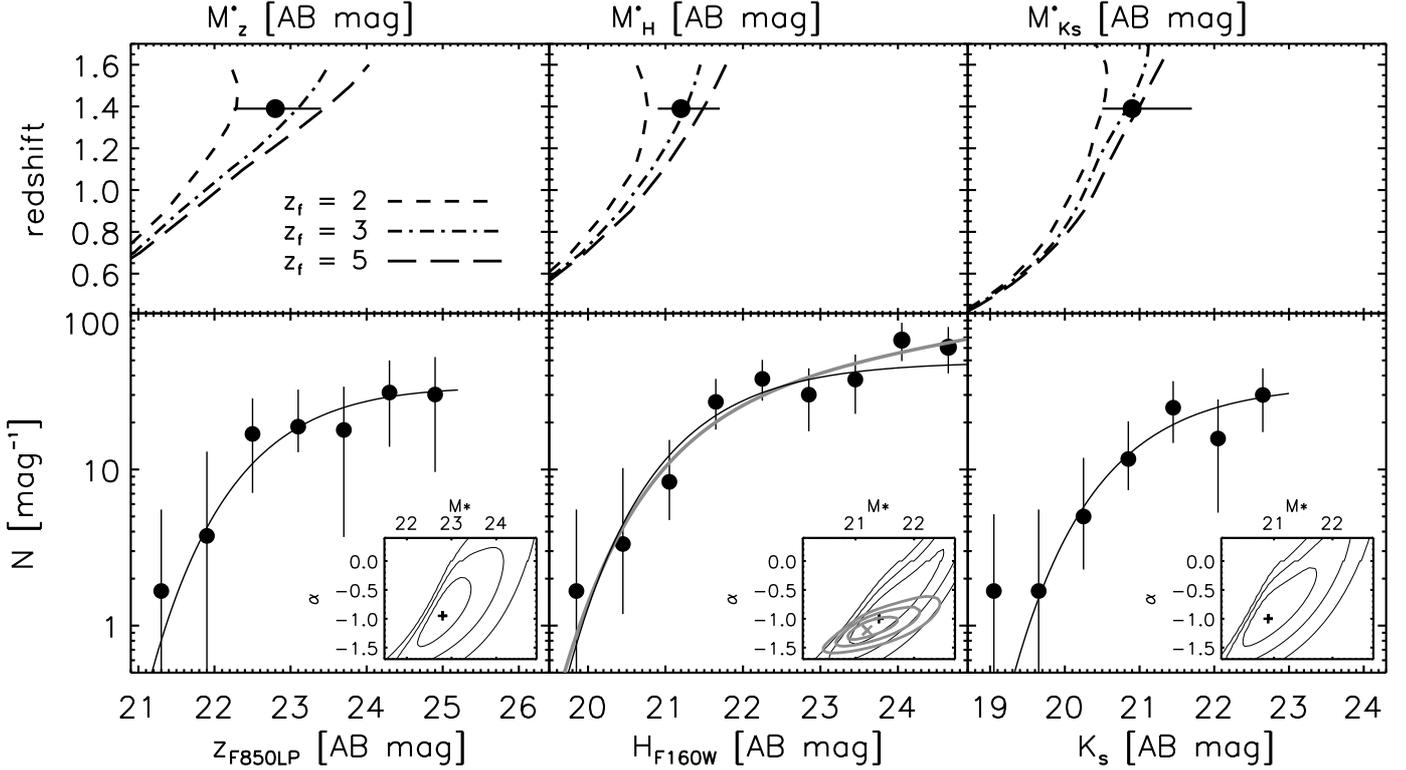}
\caption{ {\bf Bottom panels:} The luminosity function of galaxies in
the central region (within $\sim700$kpc of the cluster center) of XMMU
J2235, in the z, H and K$_{s}$ bands. Solid symbols show binned counts
(with 1$\sigma$ errors, see text for details), and black lines show
the best--fit Schechter function as determined for magnitudes brighter
than $\sim$M$^{*}$+2. Confidence levels (1,2,3$\sigma$) on the two
relevant parameters (characteristic magnitude M$^{*}$ and faint end
slope $\alpha$) are plotted in the small insets. In the central panel
(H band LF) and corresponding inset, the gray lines show the best-fit
Schechter function and confidence levels for the fit down to
$\sim$M$^{*}$+4.  The formal best--fit values and 1$\sigma$ errors for
M$^{*}$ and $\alpha$, as fitted down to $\sim$M$^{*}$+2, in the three
passbands are: M$^{*}$$_{z}$=$22.8\pm0.6$, $\alpha_{z}$=$-0.95\pm0.6$,
M$^{*}$$_{H}$=21.4$^{+1}_{-0.5}$ , $\alpha_{H}$=-1$^{+1}_{-0.4}$ (
M$^{*}$$_{H}$=21.2$^{+0.5}_{-0.3}$ ,
$\alpha_{H}$=-1.2$^{+0.2}_{-0.15}$ down to $\sim$M$^{*}$+4),
M$^{*}$$_{Ks}$=20.9$^{+0.8}_{-0.4}$,
$\alpha_{Ks}$=-1.0$^{+0.8}_{-0.5}$.  {\bf Upper panels:} The evolution
of M$^{*}$ with redshift as observed in the z, H and K$_{s}$ band,
according to \citet{kodamaearimoto} models, for three formation
redshifts $z_{f}$=2,3,5. Solid symbols show M$^{*}$ (and errors) as determined
from the LFs in the lower panels.
\label{fig:LFs}}
\end{figure*}

For the purpose of statistical subtraction of background
contamination, we also made use of publicly available photometry
acquired with the same, or very similar, instrument/filter as the
XMMU J2235 images. For the K$_{s}$ band we used VLT/ISAAC K$_{s}$ band
photometry in a $\sim60$ square arcmin region within the GOODS CDF--S
field (v2.0 release, \citet{retzlaff2009}). Only regions with a 10$\sigma$ depth
comparable or better than the XMMU J2235 K$_{s}$ image were used.  For
the H band we used the NICMOS imaging on a $\sim5$arcmin$^{2}$ region
in the Hubble Ultra Deep Field \citep{thompson2005}, while for the z
band we used ACS photometry in the GOODS (North and South) fields (v2.0
release, \citet{giavalisco2004}, Giavalisco et al. in prep.), for an
overall area of more than 200arcmin$^2$.

For all fields, point--like sources were identified and removed based
on the combination of morphological/concentration parameters estimated
by SExtractor (MAG\_AUTO, FLUX\_RADIUS, CLASS\_STAR, FWHM\_IMAGE) on
the z band image.

Luminosity functions (LFs) were determined by means of statistical
background subtraction, but at the same time taking full advantage of
the available spectroscopic and photometric redshift information. The
best purely statistical approach would avoid binning the data, and
would derive the cluster galaxy LF by simultaneously modeling the
number counts in the cluster and in a control field, as described in
detail in \citet{andreon2005}. However, this approach would not allow
us to take into account the information about cluster membership that
we derive by other means (spectroscopic and photometric
redshifts). While information on cluster membership is obviously
always relevant, its specific importance in the determination of the
LF (and its errors) depends on different factors including the
quality of such membership information and how much it contributes to
the constraints on the LF. When only one cluster field is available,
the statistics at the LF bright end is very poor; furthermore,
particularly in the case of very distant clusters, the imaging depth
can reach at most about 3-4 mag fainter than M$^{*}$. In these conditions,
it is very important to make full use of all the available
information, in order to best constrain the LF even when the
intrinsically low counts would make its purely statistical
determination relatively loose.

We thus adopted the following approach (which extends with the use of
photo-zs the approach adopted in \citet{strazzullo2006}): we first
determined the LF in each passband by subtracting the (area normalised)
counts in the control field from the counts in the cluster field. The
error on the excess counts was determined by summing in quadrature the
Poissonian errors{\footnote{Following \citet{gehrels86}, upper limits
are estimated as $N+1+(N+0.75)^{1/2}$, and lower limits as $N(1 -
\frac{1}{9N} - \frac{1}{3 \sqrt N})^{3}$.  } from both cluster and
field counts (even though it is dominated by the cluster contribution
when the cluster field is much smaller than the control field). The
error contribution coming from cosmic variance due to galaxy
clustering, as estimated according to \citet{huang1997}, is much
smaller than the Poisson error (a factor of a few percent at most) and
was neglected \citep[see also e.g.,][]{andreon2005,strazzullo2006}.

 We then took into account the spectroscopic information, imposing
that in each magnitude bin the excess counts are at least equal to the
spectroscopic members in that bin. The spectroscopic (and photo--z)
information is also considered in the determination of the final error
on the excess counts in each bin. The membership for the brightest
galaxy populations in the core of XMMU J2235 can be considered quite
well established (see also figures in section \ref{sec:stelpopcmds}
below).  In the critical (for the LF determination) range from the BCG
magnitude to M$^{*}$, the spectroscopic completeness is about
70\%. Furthermore, basically all plausible members brighter than M$^{*}$
were targeted for spectroscopic follow--up, and thus the remaining
sources without spectroscopic redshift have a photo--z far below the
cluster redshift, and well beyond 3$\sigma$ from the cluster redshift
according to the photo--z estimated rms.  Therefore, in estimating
the error on the excess counts, when all objects in a magnitude bin
have a spectroscopic redshift, or a photo--z beyond 3$\sigma$ from the
cluster redshift, we assumed that: i) the LF is basically ``membership
determined'' without the error contribution from the background
counts, ii) the lower limit on the excess counts cannot go below the
lower limit based on the spectroscopic excess counts, and iii) the
upper limit on the excess counts is limited by the upper limit on the
counts of spectroscopic members plus photo--z candidate members within
3$\sigma$ from the cluster redshift. We remind the reader that the
3$\sigma$ photo--z retained sample is estimated to be 100\% complete,
and affected by a $\gtrsim$50\% contamination, thus we expect the
upper limits based on such sample to be robust.

In Figure \ref{fig:LFs} we show the binned LFs determined in this way
in the three z, H, and K$_{s}$ passbands. The available photometry
reaches out to $\sim$M$^{*}$+2 or M$^{*}$+4 depending on the passband. For ease
of comparison, all LFs are plotted in Figure \ref{fig:LFs} over a
magnitude range of about 5 magnitudes, from $\sim$M$^{*}$-2 to
$\sim$M$^{*}$+3.5. The formal best--fit \citet{schechter1976} function
determined by $\chi^2$ minimization on the binned counts plotted, with
errors as described above, is also shown. For the H band we plot both
the LF determined for galaxies brighter than M$^{*}$+2, as for the z and
K$_{s}$ bands, and down to $\sim$M$^{*}$+4, which is only probed by the H
band. In all cases, the best--fit faint--end slope is close to flat
and the characteristic magnitude M$^{*}$ is close to the predictions of
passive evolution of the local M$^{*}$, assuming that the bulk of the
stellar populations are formed at a redshift $\sim3$. This is shown in
the upper panels of Figure \ref{fig:LFs}, showing the redshift
evolution of M$^{*}$ based on \citet{kodamaearimoto} models, in each of the
three passbands, together with the M$^{*}$ determination from the LFs in
the lower panels. The errors on M$^{*}$ are determined based on the
two--parameter confidence levels plotted in the small insets in the
lower panels.

While the K$_{s}$ band still samples the restframe near--infrared
light, which can be considered as a probe of the stellar mass, the H
and particularly z bands sample wavelengths more affected by recent or
on--going star formation activity. Nonetheless, the LFs in all three
passbands suggest a similar M$^{*}$ evolution. The flat faint--end slope,
and the measured M$^{*}$ close to passive evolution predictions of the
local cluster galaxy LF, point toward the LF bright end being similar
to that of local clusters, beside evolution of the stellar
populations. This extends to $z\sim1.4$ previous findings on the early
assembly of the massive galaxy populations in the cluster core regions 
\citep{depropris1999,nakata2001,kodama2003,toft2003,ellis2004,toft2004,andreon2006,strazzullo2006,depropris2007,muzzin2008,mancone2010}.

\section{Stellar population properties of cluster galaxies}
\label{sec:stelpopcmds}
In the following, we investigate the properties of cluster galaxies in
XMMU J2235, and of their host stellar populations, by means of direct,
simpler approaches (as the color--magnitude diagrams) and of more
detailed modeling (SED fitting with stellar population synthesis
models), using the photometric and spectroscopic measurements
described above.

\subsection{Color--magnitude diagrams}

Color--magnitude diagrams (CMDs) shown in the following refer to a
region of about $3\times3$arcmin$^2$ in the cluster center (the
maximum distance from the BCG is about 1.2Mpc). The H band image only
covers a portion of this area, corresponding to a
$\sim2.4\times2.4$arcmin$^2$ square centered on the BCG, which
translates into $\sim1200\times1200$kpc$^2$ at z=1.39.

Colors used in the following are based on resolution--matched
magnitudes measured in a 1'' aperture, which minimize the photometric
errors (and possible contamination by neighbors), especially for
fainter sources.
We note that an independent, alternative approach based on model total
magnitudes and aperture colors obtained via surface brightness fitting
of each source, gives results in very good agreement with those
obtained here (Nu\~nez et al., in preparation). We thus assume that the
systematic errors affecting MAG\_AUTO as an estimate of total
magnitude, and aperture colors obtained as described above, do not
significantly affect our results.

\begin{figure}[t]
\centering
\hspace{-.2cm}
\includegraphics[width=9cm]{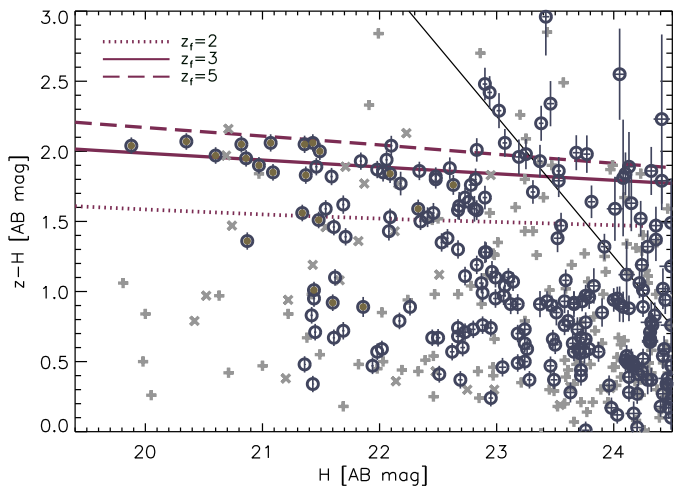}
\caption{ The color--magnitude diagram in the central region
($r\leq600$kpc) of XMMU J2235.  All sources in the field are shown,
except for point--like sources. Filled circles are spectroscopic
members and large empty circles show photo--z retained sources
($0.96<z<1.82$). Gray symbols (crosses and pluses) show spectroscopic
and photo--z interlopers, respectively. The colored lines show the
expected location of the red sequence at $z\sim1.4$ based on
\citet{kodamaearimoto} models for different formation redshifts ($z_f
$=2, 3, 5, as indicated). The slanted continuous line shows the
10$\sigma$ completeness (for the overall catalog, considering AUTO
apertures). The plotted z-H colors are derived from resolution-matched
1'' aperture magnitudes, and error bars include the estimated error
contribution from the resolution--matching procedure.  Error bars are
plotted only for retained sources.
\label{fig:CM1}}
\end{figure}

In Figure \ref{fig:CM1}, we plot the z-H vs H CMD. All objects in the
field are plotted, except for point--like sources which were removed
(see section \ref{sec:LFs}).  Spectroscopic cluster members are
plotted as filled symbols, while spectroscopic interlopers are plotted
as crosses. For consistency with \citet{lidman2008} and
\citet{rosati2009}, we define as cluster members all objects with
$1.37<z<1.41$.

In addition to excluding spectroscopic interlopers, we can further
clean the CMD in Figure \ref{fig:CM1} of obvious fore-- and background
objects excluding sources with a photo--z beyond 3$\sigma$ of the
cluster redshift (photo-z$<0.96$ or $>1.82$). These most likely
interlopers are plotted in Figure \ref{fig:CM1} as plus symbols, while
the retained sources with photo--z within 3$\sigma$ of the cluster
redshift are plotted as empty circles. The whole H$<$24.5 galaxy sample
plotted in Figure \ref{fig:CM1} contains 391 sources, out of which 205
have a photo-z within 3$\sigma$ from the cluster redshift, 21 are
spectroscopic cluster members, and 24 are spectroscopic interlopers.
As explained above, the sample selected by retaining photo-z candidate
members within 3$\sigma$ of the cluster redshift is virtually 100\%
complete, but affected by a $\geq50\%$ contamination, thus we point
out that the only purpose of the ``photo--z cleaned samples'' plotted
in the following Figures \ref{fig:CM2} and \ref{fig:CM3} is to show
more clearly the bright cluster population by removing the
contamination of obvious foreground sources.

In Figure \ref{fig:CM2}, we plot the z-H vs H and z-J vs J CMDs,
showing only spectroscopic cluster members and photo--z retained
sources. The observed z band well matches the restframe U band of the
cluster galaxies, and the J and H bands sample the restframe SED of
cluster members at $\sim$5000\AA~ and $\sim$6700\AA, respectively. The
colors shown in Figure \ref{fig:CM2} thus correspond approximately to
restframe colors U-B/U-V and U-R.  While the observed z-J color best
samples the amplitude of the 4000\AA~ break, the z-H vs H CMD benefits
from the excellent photometric accuracy and depth provided by HST
imaging. Nonetheless, the presence of a tight red sequence is evident
in both CMDs, with the zero--point and slope of the bright red
sequence in good agreement with \citet{kodamaearimoto} model
predictions for a formation redshift $z_f\sim 3$.  The intrinsic
scatter of the bright z-H red sequence (within the shaded area of
Figure \ref{fig:CM2}) is estimated to be 0.08$\pm$0.01, independent of
considering only morphological early-types or all red sequence
galaxies. This estimate of the intrinsic scatter only takes into
account photometric errors as estimated by SExtractor; considering the
error introduced by the adopted PSF matching approach could lower the
intrinsic scatter to values as low as 0.05$\pm$0.015.

We can roughly estimate the backward evolution of the XMMU J2235 red
sequence bright-end (within the shaded area in Figure \ref{fig:CM2})
with a simplistic approach. The restframe U-B scatter calculated from
synthetic colors for the red-sequence galaxies is $\sim0.06$, in
good agreement with previous estimates at similar redshifts
\citep{blakeslee2006,gobat2008,mei2009}.  Assuming that the star
formation histories of the bright red sequence galaxies may be
described by a simple exponentially declining star formation rate with
time-scale $\tau$, we evolved the $z\sim1.39$ red-sequence galaxies to
earlier epochs (see also a more detailed analysis in
\citet{gobat2008}). In general agreement with the \citet{gobat2008}
results based on the red sequence of RDCS1252 at $z\sim1.2$, we find
that beyond redshift 2 (e.g. $\sim$1Gyr earlier than the observed
epoch for XMMU J2235) a significant part of the red sequence galaxies
would no longer be classified as ``red'', and the scatter of the remaining
``red sequence'' would be substantially larger than what measured in
XMMU J2235. We note that this simplistic de-evolution of the red
sequence does not take into account the whole population of massive
galaxies in XMMU J2235 and the effects of dust extinction, thus does
not include the contribution to the red sequence by reddened
starbursts. By redshift $\sim$2.3, $\sim$ 40\% of the initial
red-sequence sources would not be ``red''; those still appearing as a
``red sequence'' would have an average age/$\tau \sim$5, thus they
would have joined the red sequence relatively recently. This is in
overall agreement with red-sequence observations at $z>2$ in
proto-clusters as well as in the field \citep[e.g.,
][]{kodama2007,kriek2008,zirm2008}.

We note that these CMDs are based on resolution-matched,
fixed--aperture photometry in a 1'' aperture, corresponding to a
physical radius of $\sim4.2$kpc, and to about 1.8 times the median
effective radius measured for the bright red--sequence
early--types. Since the resolution--matching is performed
independently on each passband image, the agreement of the formation
epochs as estimated from the z-H and z-J CMDs suggests that the
determined CMDs should not be significantly affected by systematics
introduced by the resolution--matching procedure.

The presence of a tight red sequence in XMMU J2235 has already been
reported in \citet{lidman2008}, based on the same high quality
ground--based NIR photometry used in this work. The \citet{lidman2008}
J-K$_s$ vs K$_s$ CMD was built within a distance of 900 kpc from the
cluster center, and the parameters of the red sequence were determined
based on 9 bright red--sequence galaxies within a smaller region
($r<90$kpc from the cluster center), four of which are spectroscopic
members.  Zero--point and slope of the J-K$_s$ vs K$_s$ red sequence
are in good agreement with simple stellar population (SSP) models,
reproducing the red sequence in the Coma cluster, formed at
$z\sim4$. Therefore, while the observed J-K$_s$ color at z=1.39 is
not the best choice to probe the stellar population ages, and while
even high quality ground--based imaging hardly competes with HST
imaging, especially in crowded fields, our z-H vs H color--magnitude
relation substantially confirms the results of \citet{lidman2008} on
the presence of a tight red sequence of bright, passively evolving
galaxies which formed their stars at high redshift.  In addition, we
note that the formation epoch of massive red sequence galaxies
estimated in Figure \ref{fig:CM2} by comparison with
\citeauthor{kodamaearimoto} models, are in excellent agreement with
the star-formation history derived by modeling the 
spectro-photomeric data of these galaxies \citep{rosati2009} with
\citet{bc03} models.

\begin{figure}[t]
\centering
\includegraphics[width=9cm]{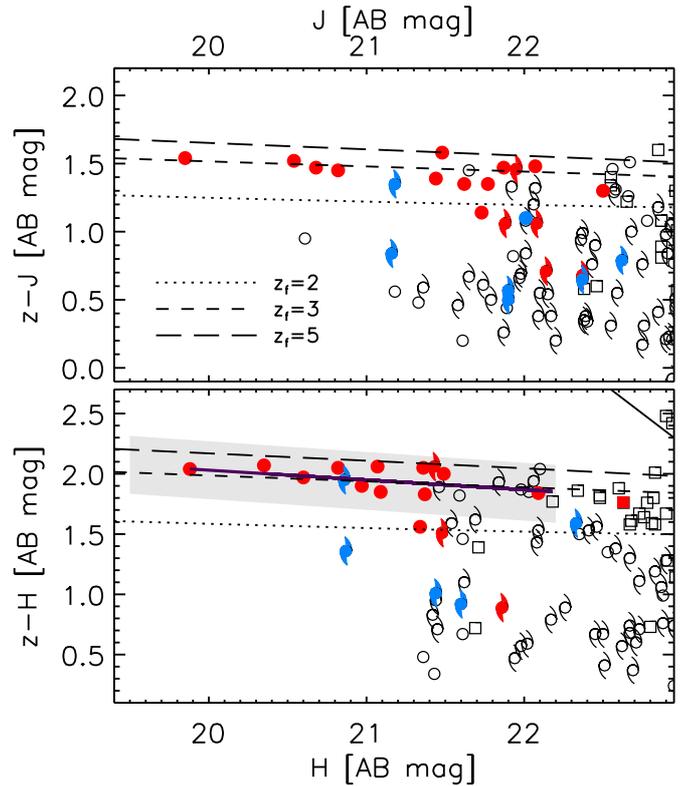}
\caption{ Color--magnitude diagrams in the central region of XMMU
J2235. Only spectroscopic members and photo--z retained sources
($0.96<z<1.82$) are shown. Filled blue and red symbols show
spectroscopically confirmed members with and without detectable [OII]
emission, respectively. The dotted, dashed and long--dashed lines show
the expected location of the red sequence based on
\citet{kodamaearimoto} models for different formation redshifts. The
dark purple line in the lower panel shows a fit to the red sequence
galaxies within the gray shaded region. The shape of all symbols are
coded according to morphological classification in the z band: circles
are galaxies with $n_{Sersic}>2$, spiral symbols are galaxies with
$n_{Sersic}<2$, squares are (mostly faint) objects for which no
reliable classification is available.
\label{fig:CM2}}
\end{figure}

In Figure \ref{fig:CM2}, spectroscopic cluster members are
color--coded according to the detection or non--detection of [OII]
emission in their spectra; however, note that due to the quality of
our spectra, [OII] equivalent widths of less than 5\AA~ cannot be
detected \citep{rosati2009}.  As noted in \citet{lidman2008} and
\citet{rosati2009}, emission line members tend to avoid the central
area of the cluster, the closest emission--line member being at a
distance of $\sim240$kpc from the BCG. We note that, even though the
spectroscopic sample is generally not complete, according to the
photo--z selection it is likely to be complete (in the central
$\sim1200\times1200$kpc$^2$ region) for sources brighter than M$^*$,
and $\geq80\%$ complete for masses larger than $10^{11}$M$_{\odot}$).

We also make use of the morphological information in the ACS z band to
broadly classify sources in early and late morphological types.  We
are mainly interested in identifying late type galaxies on the red
sequence, since these might be dusty star--forming galaxies landing on
the red sequence because of dust reddening instead of the old age of
their stars. As a result, these objects might bias our analysis and
interpretation of the red sequence. The separation of early and late
morphological types based on the Sersic index $n_{Sersic }$ alone is
known not to be very accurate \citep[e.g.,][]{blakeslee2003}, and a
multi--parametric classification should be adopted for a robust
characterization. However, in this work we only aim at classifying
galaxies in very broad morphological classes, which can be
accomplished with good statistical accuracy by using the Sersic index
alone \citep[e.g.][]{blanton2003,ravindranath2004,pannella2006}.  In
the following, we will use $n_{Sersic } >2$ as the threshold to
identify morphological early--types. With this selection, we are
likely to include basically all ellipticals, and the vast majority of
S0 galaxies, while having some contamination mainly from late--types
with a significant bulge component.  Such a threshold is a conservative
choice from the point of view of the characterization of the red
sequence sample, which will include some late morphological types. On
the other hand, when considering the morphological mix of galaxy
populations on the red sequence, we remind the reader that our
early--type sample might be affected by such contamination.

In Figure \ref{fig:CM2}, we thus plot with different symbols galaxies
for which we were able to perform a morphological analysis, according
to their Sersic index (circles for $n_{Sersic } >2$, spiral symbols
for $n_{Sersic } <2$). Most of the bright red--sequence galaxies in
the cluster core are morphologically early types according to this
criterion ($\sim90$\% down to M$^{*}$). Visual inspection of the
images of red sequence galaxies (gray shaded area in Figure
\ref{fig:CM2}) confirmed the results based on the automated
classification, in particular a dominance of early-type galaxies and a
minor contribution from late types (often hosting a prominent bulge).

All but one of the spectroscopic members classified as morphological
early types do not have detectable [OII] emission. The one exception is
a galaxy with $n_{Sersic}\sim$2.4 and intermediate colors
(z-J$\sim$1.1,z-K$_{s}\sim$1.8), located more than 900kpc away from the
cluster center. [OII] emission is detected in most of the spectroscopic
members classified as morphological late types. A fraction ($\sim30$\%) of
the (incomplete) sample of spectroscopic members classified as
morphological late types lie on or close to the red sequence.  Also, about
40\% of the spectroscopic members morphologically classified as late-types
within the studied field have no detectable [OII] emission, thus likely an [OII]
EW$<$5\AA.  All but one of these late--type galaxies without detectable
[OII] are relatively blue (as compared to the red sequence); none of
them is located in the very central area of the cluster, the closest
to the BCG is at a distance of about 150 kpc, while the others are
more than 600 kpc away.

Unfortunately, our sample of late-type spectroscopic members is
inevitably incomplete (many late-type candidate members are beyond the
spectroscopic limit).  Much larger samples will be needed to
disentangle morphological and spectro--photometric evolution of
cluster galaxies at this redshift.  We also note that [OII] emission
might also be due to AGN activity rather than star formation
\citep{yan2006}, especially for objects with a typically red
SED. Nonetheless, we also note that the only [OII] emitter actually
lying on the red sequence has a late--type morphology; its colors
would be $\sim0.2-0.3$mag bluer in a larger aperture including the
extended disk. This source is located at a distance of $\sim860$kpc
from the cluster center.  None of the red--sequence galaxies shown in
Figure \ref{fig:CM2} is detected as an X--ray point source.  We note
that the 190 ks Chandra observations of this field allow the detection
of AGN with X-ray luminosities $L_X\gtrsim 10^{43}$ erg/s in the 2-10
keV band.

\begin{figure}[t]
\centering
\includegraphics[width=9.3cm]{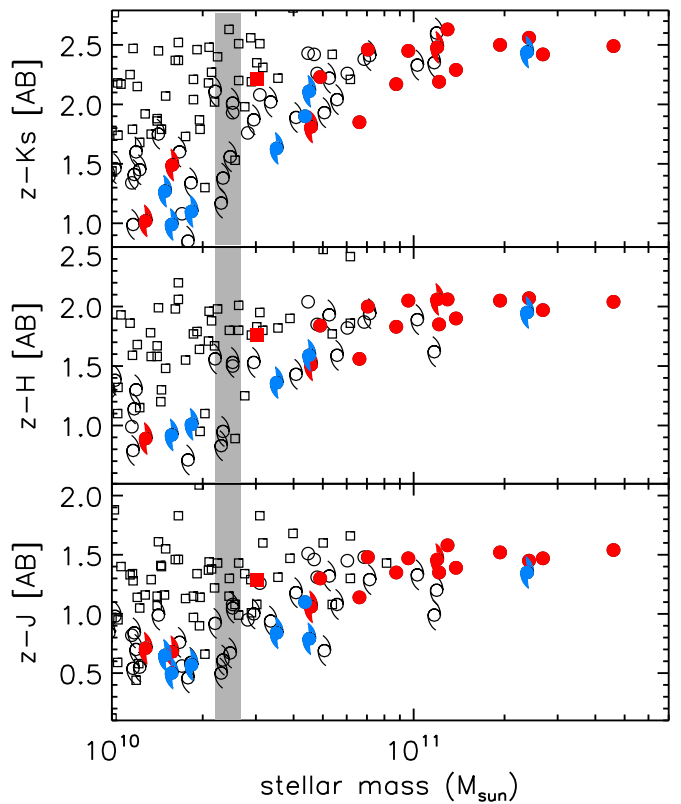}
\caption{
Color--mass diagrams in the central region of XMMU J2235. Only
spectroscopic members and photo--z retained sources ($0.96<z<1.82$)
are shown. Symbols and color coding as in Figure \ref{fig:CM2}. The
thick gray line across all panels shows the estimated mass
completeness limit (see text).
\label{fig:CM3}}
\end{figure}

\subsection{Color--mass diagrams and star formation rates}
\label{sec:cmassd}

We show in Figure \ref{fig:CM3} the color--stellar mass
diagram in the central cluster area, as in Figure \ref{fig:CM2}.
While overall similar to the color--magnitude diagram in terms
of bimodality of the galaxy populations, the color-mass diagram
yields more fundamental insight because of the disentanglement between
color, magnitude, and mass--to--light ratio.

Colors were derived from resolution--matched photometry measured in
1'' apertures, as described above. The derivation of stellar masses is
described in section \ref{sec:Mstars}, and an approximate
estimate of the mass completeness limit of this sample, derived from
the z band 10$\sigma$ completeness assuming a SSP of solar metallicity
formed at $5<z<10$, is shown by the thick gray line in Figure
\ref{fig:CM3}.

Figure \ref{fig:CM3} confirms that high mass galaxies ($\gtrsim
10^{11}$M$_{\odot}$) are already in place in the core region of XMMU
J2235, and that this high--mass population is dominated by
red--sequence galaxies, mainly of early--type morphology, generally
lacking evidence of residual star formation activity. 
The stellar mass of the BCG is close to the stellar mass of local BCGs
in massive clusters \citep[see][for a broader
picture]{collins2009,stott2010}. According to semi-analytical
predictions \citep{delucia2006}, a BCG at z$\sim$1.4 will have
typically assembled only $\sim20\%$ of the final stellar mass it will
reach at z=0. In such a model, the BCG of XMMU J2235 would thus
eventually become an object of 2.3$\cdot 10^{12}$M$_{\odot}$; while
not impossible, this would certainly put it at the very high-mass end
of local BCGs. We note however that this same model predicts a slower
evolution for BCGs of most massive, X-ray luminous clusters as
XMMU J2235.

Figure \ref{fig:CM3} clearly shows that, even at $z\sim1.4$, there are
essentially no blue galaxies in the most massive populations living in
the core of such a massive cluster.  As expected, going to lower
masses the importance of late--type galaxies (from both the
morphological and the photometric point of view) increases. It may be
interesting to note that blue, disk--dominated spectroscopic members
with [OII] emission generally lie at masses below
$\sim5\times10^{10}$M$_{\odot}$. We note again here that the
spectroscopic sample is biased against red galaxies at these masses,
because of their intrinsic faintness due to their higher
mass--to--light ratio, and to the lack of emission lines.

While keeping in mind that the rest--frame FUV light may be heavily
affected by dust attenuation, we further note here that the
rest--frame 1500\AA~ flux of massive ($>6\times10^{10}$M$_{\odot}$)
red sequence early--types is indeed consistent with these galaxies
hosting no significant star formation. After removing galaxies whose U
band photometry is likely contaminated by other sources, we stacked U
band images for eight early--types more massive than
$6\times10^{10}$M$_\odot$ (six out of these eight are spectroscopic
members). We found no significant detection and obtained a $5\sigma$
U-band upper limit of 27.5 AB mag, corresponding to a (not
dust-corrected) star formation rate (SFR) of $\sim$0.15M$_{\odot}$/yr
(for comparison, \citet{rettura2010} estimated by U-band stacking an
upper limit of $\sim$0.3M$_{\odot}$/yr for the SFR in massive
early--types in the $z\sim1.2$ cluster RDCS J1252-2927).

For comparison, the SFR of the spectroscopic members lying below the
$z_f=2$ model in Figure \ref{fig:CM2} was also estimated based on the
measured U band (rest--frame $\sim 1500$\AA) photometry. Since we know
that these are star-forming sources, we estimated a dust attenuation
correction for the $\sim 1500$\AA~rest-frame luminosity from the
restframe far-UV - near-UV color
\citep{daddi2007,salim2007,pannella2009}, which in our case in sampled
by the observed U and R bands. SFRs derived with these dust
attenuations following e.g. \citet{daddi2007}, range between 6 and
50-90$M_{\odot}$/yr depending on the assumed relation, with a median
of $\sim$20M$_{\odot}$/yr. Specific SFRs (SSFR, SFR/stellar mass)  are between 0.2 and 2-6Gyr$^{-1}$
(also depending on the specific relation used), with an average of
$\sim$1Gyr$^{-1}$. We remind the reader of the uncertainties which
affect these values due to the scatter in the assumed relations,
filter mis--match, and general difficulties in estimating correct dust
attenuation values for a given source/sample (for instance,
\citet{salim2007} note that their attenuations as a function of UV
color are systematically lower than other estimates by
e.g. \citet{meurer1999,seibert2005}). For comparison we also estimated
dust extinction for these sources with a different approach, based on
the correlation between dust attenuation and stellar mass for
star-forming galaxies, assuming that these objects follow a
correlation similar to that of field galaxies at comparable
redshift. We estimated the dust attenuation for the 1500\AA~rest-frame
luminosity based on the galaxy stellar mass according to the relation
determined for $z\sim$1.7 galaxies in \citet{pannella2009}, obtaining
an independent estimate of the SFR.  We note that since the relation
between stellar mass and dust attenuation is redshift dependent (given
the secular decline of (S)SFR), we corrected for the cosmic time
difference between $z=1.7$ and $z=1.39$ according to eq.2 in
\citet{pannella2009}. The SFRs for the blue cluster members derived in
this way range between $\sim6$M$_{\odot}$/yr and
$\sim100$M$_{\odot}$/yr, with a median of
$\sim25$M$_{\odot}$/yr. SSFRs range between $\sim$0.3Gyr$^{-1}$ and
$\sim3$Gyr$^{-1}$, with an average SSFR$\sim$1.3Gyr$^{-1}$.  Stellar
masses of these sources are in the range
1-7$\cdot10^{10}$M$_{\odot}$. Due to the observational bias, most of
the passive spectroscopic members lie at masses above this range. At
masses below $\sim 5 \cdot 10^{10}$M$_{\odot}$, with the available
spectroscopic data, redshifts can be measured only for sources with a
relevant amount of star formation (and without extreme dust
extinction). As Figure \ref{fig:CM3} shows, the star forming
spectroscopic members discussed above are among the bluest sources in
the cluster galaxy populations in their mass range.

Finally, Figure \ref{fig:panorama} summarizes the information
concerning stellar masses, (spectro-)photometric and morphological
properties of 2$\sigma$ photo-z retained candidate cluster members
together with their projected distance from the cluster centre. Recent
work on high redshift clusters at $z\sim1.45$ and 1.62, corresponding
to cosmic times about 0.2Gyr and 0.6Gyr earlier than XMMU J2235,
suggested that a significant star formation activity might be
occurring in the very central regions \citep[$r<250-500$kpc,][]{hilton2010,tran2010} of
clusters in a less advanced evolutionary stage.
However, as Figure \ref{fig:panorama} shows, star formation appears to
be effectively quenched in the core of a more evolved, massive X-ray
luminous cluster as XMMU J2235. The relevance of very reddened star
formation contaminating the cluster red sequence could be addressed
with future observations (NIR spectroscopy, Herschel IR photometry),
however, even though the rest-frame NUV (U band) photometry might be
affected by dust extinction, the early-type morphological appearance
and spectral features of the central massive red galaxies suggest
that these are more likely passive early-types than highly attenuated
starbursts.

\begin{figure}[t!]
\centering
\includegraphics[width=9cm]{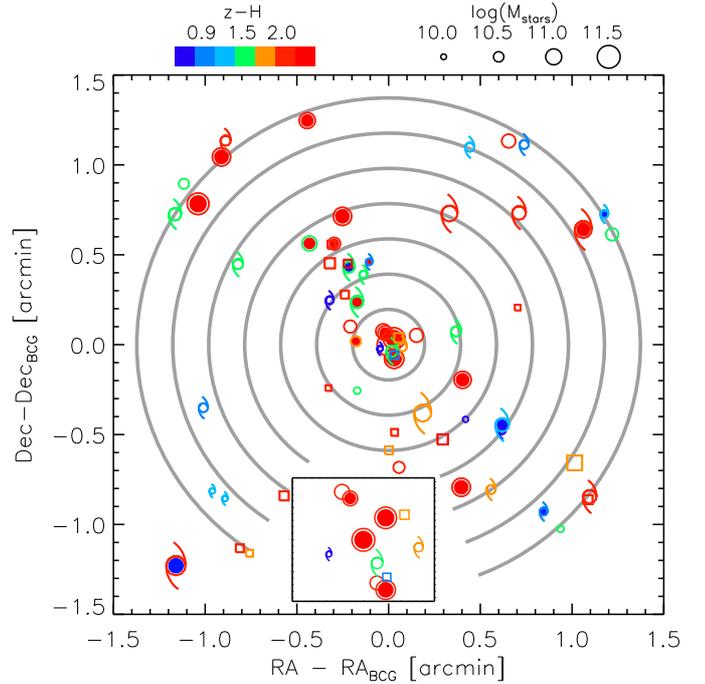}
\caption{ The projected distribution around the cluster BCG of
2-$\sigma$ retained photo-z candidate members (spectroscopic
interlopers are not plotted). Symbol shape is coded as in Figures
\ref{fig:CM2} and \ref{fig:CM3}, external color coding reflect the z-H
color while symbol size scales with stellar mass (as indicated). Color
filling, for spectroscopic members, is blue for galaxies with detected
[OII] emission, red otherwise. The small inset at the bottom of the plot
shows a larger view of the central $\sim0.2\times0.2$arcmin$^2$
($\sim 98\times98$kpc$^2$). Gray lines show radii of 100 to 700 kpc
(with a step of 100 kpc).
\label{fig:panorama}}
\end{figure}

\section{Stellar masses and sizes of massive cluster early--types}

The above analysis in the central regions of a particularly massive,
distant X--ray luminous cluster shows that the high--mass population
of cluster galaxies primarily of early--type morphology is
substantially in place at epochs earlier than redshift one, in
relation to both the formation of the bulk of their stars (at $z>2$)
and of the assembly of the bulk of their stellar mass. This points
toward little evolution occurring in massive galaxy populations in the
most dense environments over the last 9 billion years. On the other
hand, it is interesting to remember that the structural properties of
such massive objects might be different from those of similar galaxies
in the local Universe. As pointed out in several studies
\citep[e.g.,][and references therein]{daddi2005,
trujillo2006a,trujillo2006b,
trujillo2007,longhetti2007,cimatti2008,buitrago2008,vanderwel2008,damjanov2009,rettura2010,williams2010,taylor2010},
massive early--type galaxies at high redshift appear to be (on average) more
compact than their local counterparts, their size being smaller than
that of local early--types of comparable mass (but see also results
and discussion in e.g. \citet{mancini2009,onodera2010,
valentinuzzi2010,saracco2010}, and references therein).  Indeed, as
shown in Figure \ref{fig:masssize}, also the massive early--type
galaxies in the core of XMMU J2235 appear to be generally smaller than
those in the local Universe.

Figure \ref{fig:masssize} shows the stellar mass against the
(circularized) effective radius\footnote{Six (one) sources out of
the 15 plotted in Figure \ref{fig:masssize} have a formal uncertainty
on their size lower than 15\% (10\%). Keeping into account results from
simulations or multiple independent fitting
\citep[e.g.][]{pignatelli2006,buitrago2008,pannella2009b} on the
typical uncertainty on the galaxy size ($\sim$10-15\%) achievable by
surface brightness fitting for sources of S/N similar to those used
here, we increase the error bar for these six sources to reflect an
uncertainty of 15\%.} as measured from the surface brightness fit in
the z band (section \ref{sec:galfit}), for bright
early--type\footnote{Galaxies in this sample were classified as
``early-types'' based on their Sersic index larger than 2, as
discussed above. We note that out of 15 galaxies plotted in figure
\ref{fig:masssize}, only two (with Sersic indices 2.3 and 2.4) would
not have entered the ``early-type sample'' if using a Sersic index
threshold of 2.5.} red sequence galaxies (within the gray shaded area
in Figure \ref{fig:CM2}). Most of this sample is made of spectroscopic
cluster members. For comparison we show the local stellar mass vs size
relation as measured for SDSS early--type galaxies by
\citet{shen2003}, which is commonly used as the local reference. We
note that the size used by \citet{shen2003} was measured in the z
band, while in our case galaxy sizes are measured in the observed z
band, corresponding to the restframe U band, which might imply issues
due to morphological K--corrections. However, sizes measured in bluer
bands are expected to be larger than sizes measured in redder bands
\citep[e.g.,][]{barden2005,mcintosh2005,trujillo2007}.  Therefore our
galaxy sizes, if measured in the restframe z band would be, if
anything, smaller than those plotted in Figure \ref{fig:masssize}, as
it is indeed suggested for this very sample by the comparison of
surface brightness fitting results in the observed z and K$_{s}$
passbands (Nu\~nez et al., in preparation). Therefore, taken at face
value, the comparison of our early--type sample with the local
\citet{shen2003} reference implies that the size of massive
early-types in XMMU J2235 is on average about $\sim$50\% of what
expected based on the local relation (median
r/r$_{z=0}$=0.46$\pm$0.08). We note that the most massive object in
Figure \ref{fig:masssize} is the cluster BCG; while the size of this
galaxy is quite large and it lies on the local relation, we note that
BCGs tend to be larger than similarly massive (non-BCG) galaxies
\citep[e.g.,][]{vonderlinden2007}.

\begin{figure}[b!]
\centering
\includegraphics[width=9cm]{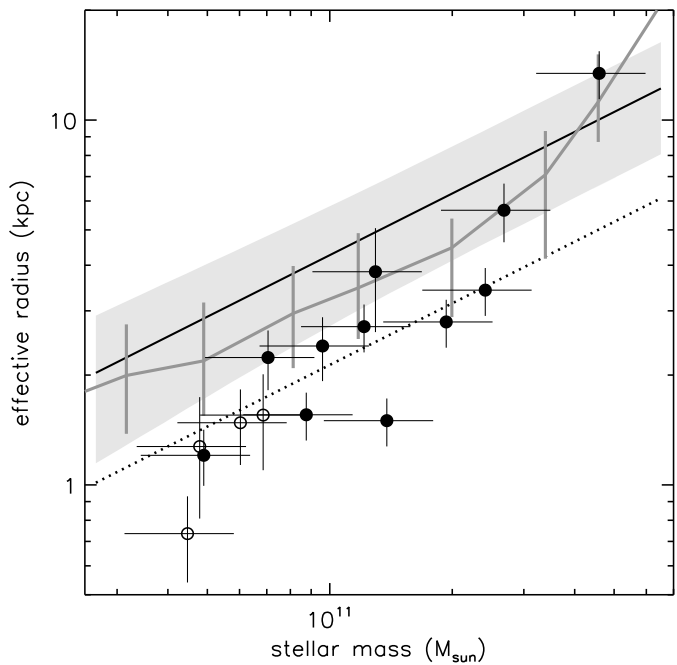}
\caption{ The stellar mass vs size relation for massive red--sequence
early--type galaxies in the core of XMMU J2235. Filled symbols are
spectroscopic members, empty symbols are sources with photo--z within
3$\sigma$ of the cluster redshift. The black line and shaded area show
the \citet{shen2003} determination of the local stellar mass vs size
relation and its scatter (1$\sigma$), based on a sample of SDSS
early--types. The dotted line shows the \citet{shen2003} relation
shifted by a factor 2 in size. The gray line with errorbars shows the
local mass vs size relation determined by \citet{valentinuzzi2010}
for a sample of nearby cluster galaxies. 
\label{fig:masssize}}
\end{figure}

The smaller sizes of XMMU J2235 massive early-types compared to local
early--types suggest a scenario where the evolution of these massive
red--sequence galaxies is not actually fully completed at early
epochs, as would be suggested by the color--magnitude and color--mass
relations, as well as the LF evolution. This leaves room for processes
such as minor (and likely fairly dry) merging events to be relevant at
later epochs to shape the final structural properties of these
galaxies, without substantially altering their stellar masses and
stellar populations (see e.g. \citet{hopkins2010} and references therein;
see also discussion in \citet{nipoti2009}).

On the other hand, we should note that the comparison of the stellar
mass vs size relation with a local sample, as shown in Figure
\ref{fig:masssize}, may be affected by systematics in the
determination of stellar masses at different redshifts, observational
biases in the determination of galaxy sizes
\citep[][]{pannella2009b,mancini2009}, and a possible mismatch between
the observed high redshift sample and the local reference, which are
discussed in more detail below.

In Figure \ref{fig:masssize} we also plot the local stellar mass
vs. size relation for a sample of nearby cluster galaxies as measured
by \citet{valentinuzzi2010}. While consistent at 1$\sigma$ with the
\citet{shen2003} relation, the \citet{valentinuzzi2010} determination is
systematically offset (possibly due to an offset with respect to the
stellar mass estimates used in \citet{shen2003}, rather than because
of environmental effects, see discussion in the original paper).
Using the \citet{valentinuzzi2010} relation as the local reference
thus reduces the difference between our sample and local massive
(cluster) early-types (median r/r$_{z=0}$=0.64$\pm$0.08).

In addition, \citet{maraston2005} has pointed out how a different
treatment of the thermally pulsing asymptotic giant branch (TP-AGB)
phase of stellar evolution may introduce additional (to those
mentioned in section \ref{sec:cmassd}) systematics in the stellar
masses, whose significance depends on the age of the stellar
populations. This may be particularly relevant when comparing
early-type galaxies at different redshifts, because while local
early-types are too old for TP-AGB phase making a significant
difference, at higher redshifts they inevitably become closer to
hosting the intermediate age stellar populations for which a different
TP-AGB phase treatment may have an important effect. For the sample
relevant to Figure \ref{fig:masssize}, stellar masses estimated with
the \citet{maraston2005} models are lower by about 0.1dex than the
masses plotted, estimated with \citet{bc03}.  Using these lower
masses, the size of galaxies in Figure \ref{fig:masssize} are
$\sim50$\% and $\sim70$\% of those of early--types at z=0, based on
the local relations by \citet{shen2003} and
\citet{valentinuzzi2010} (r/r$_{z=0}$=0.5$\pm$0.1 and 0.72$\pm$0.1,
respectively).  We note that if we use the updated version of the
\citet{bc03} models (often referred to as CB07), incorporating an
improved prescription for the treatment of the TP-AGB phase, stellar
masses for the Figure \ref{fig:masssize} sample are lower (in median)
by 0.03dex; CB07 mass estimates for this sample range from almost 30\%
lower to $\sim$10\% higher compared to \citet{bc03}.

A combination of systematics in stellar masses and mismatch with the
local reference relation might thus shift the (median) difference in
size of XMMU J2235 and local early-types between a factor slightly
over 2 and a factor 1.4.  

We note that other effects might also systematically bias the stellar
masses and sizes used above. Again, we recall that our sizes were
estimated in the restframe U band, while we are comparing with local
galaxy sizes measured in the z \citep{shen2003} and V
\citep{valentinuzzi2010} passbands. Based on previous work
\citep{mcintosh2005,barden2005,trujillo2007,szomoru2010}, correcting
our sizes to the restframe z/V bands might reduce our sizes by a
factor of up to 30\%, and thus increase the median difference in size
with local counterparts to up to a factor $\sim3$ (using \citet{bc03}
stellar masses and the \citet{shen2003} local reference).

We also remind the reader that the stellar masses used here were
  derived assuming a fixed, solar metallicity. Should the metallicity
  of the galaxies in Figure \ref{fig:masssize} be lower than solar,
  their stellar masses would be higher, by a factor about 15\% and 2\%
  for metallicites of 0.004 and 0.008, respectively. If their
  metallicity is supersolar, their stellar masses would be lower, by a
  factor $\sim$30\% for a metallicity of 0.05.

Finally, we note that the sizes we used were estimated based on a
Sersic profile fit with varying n$_{Sersic}$. Therefore, for the
sample of early types plotted in Figure \ref{fig:masssize},
n$_{Sersic}$ is not fixed to 4, but ranges between $\sim2.5-6$ with an
average n$_{Sersic}\sim3.5$. Because of the correlation between the
estimated n$_{Sersic}$ and size, an object with a n$_{Sersic}$=4
profile which is fitted with a lower Sersic index (possibly due to the
faintness of the source) will also have its size biased to a lower
value. In order to estimate the relevance of such an effect, we fitted
all sources in Figure \ref{fig:masssize} with n$_{Sersic}$ fixed to
4. This produces a very mild difference (on average $\lesssim$15\%) in
the estimated sizes, and overall negligible difference ($\lesssim$5\%)
in the evolution factors quoted above.

Although the above results imply an evolution of the stellar mass vs
      size relation by a factor of about 1.4 to more than 2, this
      evolution does not necessarily imply that the invidual galaxies
      in our early-type sample will evolve by this factor by redshift
      zero. Several studies showed how the population of early-type
      galaxies is increased, as a function of cosmic time, by the
      evolution of late-type, star-forming galaxies into quiescent
      early-types \citep[e.g., among many others,][and references
      therein]{bell2004, pannella2006, faber2007,brown2007,franx2008}. It is
      thus very likely that a significant fraction of early-types at
      z=0 were still forming stars at z=1.4 (or did not yet have an
      early-type morphology), and were added to the local "reference"
      early-type sample at later cosmic times.  Together with the
      present-day correlation (at fixed mass) between galaxy size and
      age of its host stellar populations, this may indeed suggest
      that the comparison of high-redshift early-type samples with
      local samples might artificially produce a "size evolution"
      signature, because the high-redshift sample does not account for
      the whole progenitor population of the local sample (see
      discussion in e.g.
      \citet{franx2008,vanderwel2009,valentinuzzi2010,bernardi2010,williams2010}).
      According to \citet{vanderwel2009} (their Figure 5) and
      \citet{valentinuzzi2010} (their Figure 4), the systematic
      exclusion from the high-redshift sample of the progenitors of
      younger local early-types, which tend to have larger sizes,
      might produce a difference in size of a factor as large as 1.3 at
      the redshift of XMMU J2235.

\begin{figure}[b!]
\centering
\includegraphics[width=9cm]{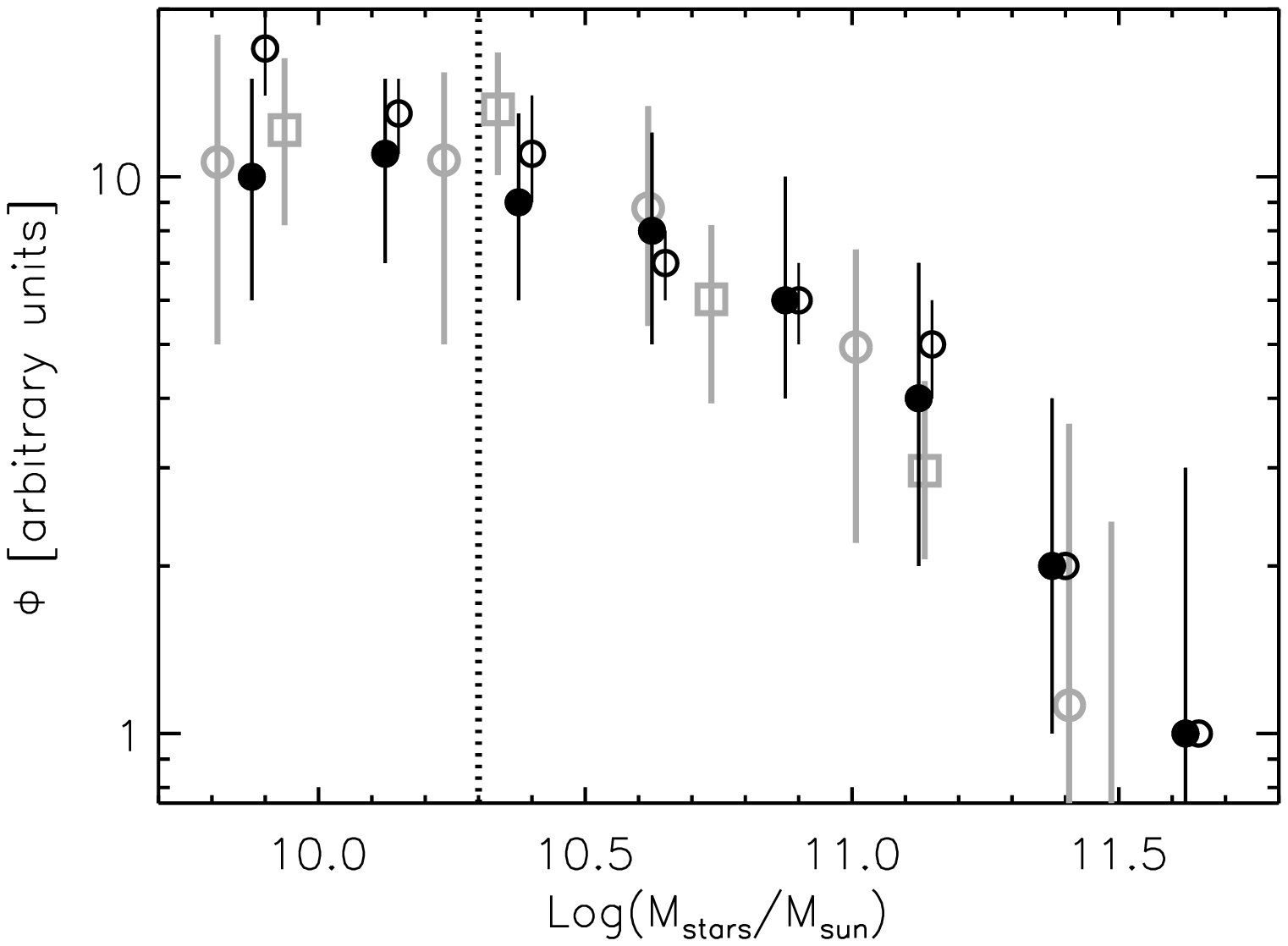}
\caption{
The galaxy stellar mass function in the central region of XMMU J2235,
as derived from the K$_s$-band LF (solid symbols). Empty black circles
show the stellar mass function derived from individual SED-estimated
stellar masses of spectroscopic and photo-z retained members (see text
for details). Gray symbols show previous determinations
\citep{kodama2003,strazzullo2006} at $z\sim1$ (squares) and $z\sim1.2$
(circles). All mass functions are arbitrarily rescaled. The dotted
line shows the mass completeness limit as estimated from the K$_s$
image 10$\sigma$ completeness.
\label{fig:MF}}
\end{figure}

A definite conclusion on the actual relevance of size evolution for
  our sample of massive cluster early-types is thus precluded by the
  possible biases mentioned above. It remains clear that the
  significant range of galaxy sizes at all redshifts requires much
  larger samples in order to draw a conclusive picture of the mass and
  environmental dependence of galaxy size evolution.

Finally, for the sake of completeness, we note that by comparing the
sizes of our spectroscopic members of {\it late-type} morphology with
the typical size of late-type galaxies in the nearby Universe
(\citet{shen2003}, as well as \citet{maltby2010} specifically in
``cluster'' and ``cluster core'' environments), we cannot find
evidence of significant size evolution at fixed stellar mass for
disk-dominated galaxies. The formal difference we find by comparing
our late-type sizes at masses $10^{10}-10^{11}$M$_{\odot}$ with
cluster galaxies at z$\sim$0.2 \citep{maltby2010} would be up to
$\sim$15\% (for our standard \citet{bc03} stellar masses), but given
the large uncertainties inherent in the comparison (as discussed
above), and the fact that our spectroscopic late-type sample suitable
for this comparison is small ($\lesssim10$ galaxies) and incomplete,
we cannot derive any definite conclusion.

\section{The stellar mass function}
\label{sec:MF}

Since the observed K$_{s}$-band light probes the rest-frame z band
of galaxies at the cluster redshift, the K$_s$-band LF derived
in section \ref{sec:LFs} can be used to estimate the stellar mass
function of galaxies in the central region ($r\lesssim r_{500}$) of
XMMU J2235.

We note that, while a galaxy NIR luminosity has a milder sensitivity
to recent star formation and dust attenuation as compared to
luminosity at shorter wavelengths, stellar masses derived from
the NIR luminosity alone may still suffer from systematics of up to a
factor $\sim$2 depending on different mass-to-light (M/L) ratios of
different galaxies.  Based on stellar masses of sources with
photo-zs within $2\sigma$ of the cluster redshift, we calibrated the
stellar mass vs observed K$_s$-band luminosity relation finding a
scatter of $\sim0.3$dex for the whole sample and of $\sim0.15$dex when
considering blue and red galaxies separately (using a color threshold
of z-H$\simeq$1.5).

We therefore applied a statistical correction for different M/L ratios
of cluster galaxies in the K$_s$-band selected sample, and derived the
mass function and its uncertainties as follows.  Firstly, we estimated
the fraction of red and blue cluster galaxies as a function of
K$_s$-band luminosity based on the $2\sigma$ photo-z retained sample,
based on both the z-H and z-J colors, yielding consistent
results. Because of the $\sim$100\% completeness and $\sim$50\%
contamination of this sample, we may expect that the estimated
relative contributions of red and blue cluster galaxies are possibly
biased toward a higher blue fraction, since the interlopers
contaminating our sample will be, on average, preferentially bluer
than cluster galaxies. This would result in a possibly lower M/L ratio
used to statistically convert the K$_s$-band light to stellar mass
(and thus lower masses). 

Once the relative contributions of red and blue galaxies, and the
typical M/L ratios of these two populations were determined, the K-band
LF was translated into a stellar mass function by means of 1000
realizations taking into account the error on the LF (as plotted in
Figure \ref{fig:LFs}), the estimated scatter in the M/L ratios of red
and blue galaxies, and the error in the red/blue galaxy fractions.

The derived stellar mass function is shown in Figure
\ref{fig:MF}. Solid black symbols and errorbars show the median and
16-84$th$ percentiles of the distribution obtained through the 1000
realizations. For comparison, we also show (empty black symbols) the
stellar mass function that is directly obtained from the individual
SED-estimated stellar masses of the spectroscopic members plus photo-z
retained sources randomly sampled assuming a $\sim$50\% contamination
by interlopers. These two largely independent determinations of the
stellar mass function in the central region of XMMU J2235 are
perfectly consistent. The shape of the mass function determined here
is in very good agreement with (arbitrarily rescaled) previous
measurements in clusters at $z\sim1$ and $z\sim1.2$ \citep[][both
scaled to a Kroupa IMF]{kodama2003,strazzullo2006}, once again
suggesting an early assembly of massive galaxies at least in overdense
environments, with the high-mass end population essentially in place
at one third of the Hubble time.

The area considered here lies within $r<$90''=765kpc, which is very
close to the estimated $r_{500}$ for XMMU J2235
\citep[$r_{500}\sim$0.75Mpc,][]{rosati2009}.  The total projected
stellar mass within this area, inferred from integrating the K$_{s}$
band LF and converting the total K-band luminosity
(L$_{K,500,proj}$=$\Phi^* L^* \Gamma$(2+$\alpha$)=11$\pm
2~10^{12}$L$_{\odot}$) to stellar mass, is
M$_{stars,500,proj}$=6$^{+6}_{-3}\cdot 10^{12}$M$_{\odot}$, with
errors including the uncertainties on the LF parameters and the
scatter in the galaxy M/L ratios. For comparison, the stellar mass in
galaxies more massive than $\sim10^{10}$M$_{\odot}$, statistically
determined from the photo-z retained sample within the same area (see
mass function in Figure \ref{fig:MF}), is $\sim3\cdot10^{12}$M$_{\odot}$.

Following \citet{rosati2009} the total projected mass within this
radius is M$_{tot,500,proj}= 7\pm 2 \cdot 10^{14}$M$_{\odot}$. The
ratio of total mass to K-band luminosity in the projected area within
$r_{500}$ is thus M$_{tot,500,proj}$/L$_{K,500,proj}$=60$\pm$20
M$_{\odot}$/L$_{\odot}$, while the stellar mass fraction is
M$_{stars,500,proj}$/M$_{tot,500,proj}$=0.009$^{+0.01}_{-0.007}$, in
close agreement with determinations for massive clusters at lower
redshifts \citep[e.g.,][]{lin2003,ettori2009,giodini2009,andreon2010}.

\section{Summary}

We presented a study of galaxy populations in the core of the massive
cluster XMMU J2235 at z=1.39, making use of high quality VLT and HST
multi--wavelength photometry in seven passbands from U to K$_{s}$,
sampling the rest--frame SED of cluster members from the NUV to the
NIR.

We derived luminosity functions in the z, H, and K$_{s}$ bands,
approximately corresponding to the rest--frame U, R, and z band.
These extend the study of the galaxy luminosity function in massive
X-ray luminous clusters beyond $z\sim 1.3$, and are among the deepest
determinations at $z>1$.  All three determinations of the luminosity
function, probing down to $\sim$M$^{*}$+2 or M$^{*}$+4 depending on
the passband, are consistent with having a flat faint--end slope and a
characteristic magnitude M$^*$ close to passive evolution predictions
of the M$^*$ of local massive clusters with a formation redshift
$z>2$.

The color--magnitude and color--mass diagrams in the core region of
XMMU J2235 show evidence of a tight (intrinsic scatter $\lesssim$0.08)
red sequence of massive galaxies, with overall old stellar populations
($z_f >2$), no evidence of significant on--going star formation
(SFR$<0.2$M$_{\odot}$/yr), and generally early--type morphology. The
spectroscopic sample is essentially complete at the high--mass end,
and strongly suggests that the most massive cluster galaxy populations
in the core of this cluster are already dominated by early--type
galaxies, both in terms of galaxy structure and of overall
spectro--photometric properties.

Star formation appears effectively quenched at masses higher than
$\sim 6\cdot 10^{10}$M$_{\odot}$. Also, active star formation is
suppressed in the very central regions, with all spectroscopically
confirmed star forming cluster members located at $r\gtrsim250$kpc
from the BCG.

In agreement with previous work at lower redshifts, these data point
toward an early assembly of massive cluster galaxies, not only in
terms of the formation of their stars, as suggested by their
broad--band SEDs, as well as by their spectra \citep{rosati2009}, but
also of the assembly of their stellar mass, as suggested by the
SED--derived stellar masses of individual sources, by the
close--to--passive evolution of the luminosity function bright-end,
and by the galaxy stellar mass function. As an additional evidence
that XMMU J2235 is in a very advanced evolutionary stage already at
this redshift, we estimate a stellar mass fraction within $r_{500}$ of
about 1\%, similar to local massive clusters.

On the other hand, comparing the size of these massive red--sequence
galaxies to the size of similarly massive local early--types, suggests
a possible size evolution of up to a factor $\sim2$.  This implies
that, in spite of the overall early assembly of these sources, room is
left for processes like minor (and likely dry) merging to shape the
structural properties of these objects to resemble those of their
local counterparts, without substantially affecting their stellar mass
or galaxy populations. We discussed how the role of possible
systematics makes it still difficult to draw firm conclusions on the
magnitude of such a size evolution, and on whether this evolution
might be different from the one estimated in field galaxy samples over
the same redshift range. Much larger samples of high-redshift
(cluster) early-type galaxies are needed to derive a conclusive
picture of the mass and environmental dependence of their size
evolution. 

Future observations at IR, mm and radio wavelengths, probing 
dust-unbiased star formation and the cold gas component, will enhance our 
understanding of the mass growth, star formation, and structural evolution 
of high-redshift cluster galaxies.

\begin{acknowledgements}
We are grateful to T. Kodama for providing us with results from his
elliptical galaxy evolution models. We thank the anonymous referee for
a constructive report which helped us to improve the presentation of
this work. VS acknowledges support under the ESO visitor program in
Garching during the completion of this work. VS and MP acknowledge
support from the Max-Planck Society and the Alexander von Humboldt
Foundation, and from NASA through Jet Propulsion Laboratory contract
No.1289215.  The National Radio Astronomy Observatory is a facility of
the National Science Foundation operated under cooperative agreement
by Associated Universities, Inc. PR acknowledges support by the DFG
cluster of excellence Origin and Structure of the Universe. Financial
support for this work was partly provided by NASA through program
GO-10496 from the Space Telescope Science Institute, which is operated
by AURA, Inc., under NASA contract NAS 5-26555. This work was also
supported in part by the Director, Office of Science, Office of High
Energy and Nuclear Physics, of the U.S. Department of Energy under
Contract No. AC02-05CH11231.

\end{acknowledgements}

\bibliographystyle{aa}

\bibliography{15251}

\end{document}